\documentclass[aps,pre,twocolumn,showpacs,graphicx]{revtex4}
\usepackage{latexsym}
\usepackage{amssymb}

\usepackage{graphicx}% Include figure files
\usepackage{dcolumn}% Align table columns on decimal point
\usepackage{bm}% bold math
\usepackage{setspace}
\begin{document}
\bibliographystyle{apsrev}

\newcommand{\la}{\left\langle}
\newcommand{\ra}{\right\rangle}
\newcommand{\Bioc}{{\it Biochemistry~}}
\newcommand{\Biom}{{\it Biomacromolecules~}}
\newcommand{\Biop}{{\it Biopolymer~}}
\newcommand{\EPJ}{{\it Eur.~Phys.~J.~}}
\newcommand{\EPL}{{\it Europhys.~Lett.~}}
\newcommand{\JCIS}{{\it J. Coll. Int. Sci.~}}
\newcommand{\JPC}{{\it J. Phys. Chem.~}}
\newcommand{\JPCM}{{\it J. Phys.: Condens. Matter~}}
\newcommand{\JCP}{{\it J. Chem. Phys.~}}
\newcommand{\Macro}{{\it Macromol.~}}
\newcommand{\MCS}{{\it Macromol. Chem. Phys.~}}
\newcommand{\MP}{{\it Mol.~Phys.~}}
\newcommand{\PR}{{\it Phys.~Rev.~}}
\newcommand{\PRL}{{\it Phys.~Rev.~Lett.~}}

\title{Phase Separation of Charge-Stabilized Colloids: \\
A Gibbs Ensemble Monte Carlo Simulation Study}

\author{Ben Lu and Alan R. Denton\footnote{Corresponding author:
{\tt alan.denton@ndsu.edu}}}
\affiliation{Department of Physics, North Dakota State University,
Fargo, North Dakota, 58105-5566}

\date{\today}

\begin{abstract}
Fluid phase behavior of charge-stabilized colloidal suspensions is explored 
by applying a new variant of the Gibbs ensemble Monte Carlo simulation method 
to a coarse-grained one-component model with implicit microions and solvent.  
The simulations take as input linear-response approximations for effective 
electrostatic interactions -- hard-sphere-Yukawa pair potential and one-body 
volume energy.  The conventional Gibbs ensemble trial moves are supplemented 
by exchange of (implicit) salt between coexisting phases, with acceptance 
probabilities influenced by the state dependence of the effective interactions.
Compared with large-scale simulations of the primitive model, with explicit 
microions, our computationally practical simulations of the one-component model
closely match the pressures and pair distribution functions at moderate 
electrostatic couplings.  For macroion valences and couplings within the 
linear-response regime, deionized aqueous suspensions with monovalent microions
exhibit separation into macroion-rich and macroion-poor fluid phases below a 
critical salt concentration.  The resulting pressures and phase diagrams are 
in excellent agreement with predictions of a variational free energy theory 
based on the same model.  
\end{abstract}


\maketitle

\section{Introduction}\label{Introduction}
Charge-stabilized colloidal suspensions~\cite{pusey} containing monovalent 
microions reportedly can display unusual thermodynamic behavior when 
strongly deionized.  Puzzling experimental observations include liquid-vapor 
coexistence~\cite{tata92}, stable voids~\cite{ise94,ise96,tata97,ise99}, 
contracted crystal lattices~\cite{matsuoka,ise99,groehn00}, and metastable 
crystallites~\cite{grier97}.  Such phenomena reveal an extraordinary cohesion 
between like-charged macroions that appears inconsistent with the 
purely repulsive electrostatic pair interactions predicted by the classic 
theory of Derjaguin, Landau, Verwey, and Overbeek (DLVO)~\cite{DL,VO}.
Failure of the DLVO theory to account for anomalous phase behavior of 
deionized suspensions has prompted many theoretical and simulation studies.

% Theory and simulation: 
Predictions of a spinodal instability in deionized charged colloids follow 
from classical density-functional~\cite{vRH97,vRDH99,vRE99,zoetekouw_pre06}, 
extended Debye-H\"uckel~\cite{warren00,chan85,chan01}, and 
linear-response~\cite{silbert91,denton99,denton00,denton04,denton06} theories 
of a coarse-grained one-component model.  The predicted phase separation is 
driven by the state dependence of the effective electrostatic interactions,
including a one-body volume energy~\cite{hansen-lowen00,belloni00,levin02}.
Such predictions have been challenged on grounds that underlying 
linearization approximations may fail to describe nonlinear microion 
screening~\cite{vongrunberg01,deserno02,tamashiro03} and neglect strong 
counterion association that may renormalize the effective macroion 
charge~\cite{levin98,levin01,levin03,levin04,zoetekouw_prl06}.  The debate 
is somewhat complicated, however, by the proximity of the unstable parameter 
regime to the threshold for significant nonlinearity and charge renormalization.

Some simulations of the primitive model~\cite{linse00,hynninen05}, with explicit
microions interacting via long-ranged Coulomb potentials, exhibit clustering
of macroions at strong electrostatic couplings.  Such computationally intensive
simulations become increasingly demanding, however, upon approaching the size 
and charge asymmetries required to directly test predictions, even when
sophisticated cluster moves are included~\cite{luijten04}.  Therefore, 
reconciling theories, simulations, and experiments to clarify the 
phase behavior of deionized charged colloids calls for novel simulation 
methods tailored to mesoscale models.

% Purpose, conclusions, and outline:
The main purpose of the present work is to propose a new variant of the Gibbs
ensemble Monte Carlo method suited to modeling density-dependent effective 
electrostatic interactions.  As a demonstration, we apply the method to
deionized charged colloids to test predictions of phase instability.  After 
first defining the model system and one-component mapping in Sec.~\ref{Models}, 
we briefly summarize the linear-response theory of effective interactions and 
a variational free energy theory in Sec.~\ref{Theory}.  The Monte Carlo 
algorithm is next outlined in Sec.~\ref{Simulations}.  Simulation results 
are presented in Sec.~\ref{Results}, with diagnostic details deferred to the 
Appendix.  Comparisons with theory and primitive model simulations confirm 
previous predictions and illustrate computational advantages and limitations 
of the one-component model.  Finally, Sec.~\ref{Conclusions} summarizes 
our conclusions.

\section{Model}\label{Models of Charged Colloids}\label{Models}

\subsection{Primitive Model}\label{Primitive}
\vspace*{-0.2cm}
As underlying microscopic model, we adopt the primitive model of charged 
colloids~\cite{HM}: macroions and microions dispersed in a continuum solvent 
of dielectric constant $\epsilon$ in a closed volume $V$.  The macroions are 
modeled as charged hard spheres, monodisperse in radius $a$ and effective 
valence $Z$ (charge $-Ze$), and the microions (counterions and salt ions) 
as point charges of valence $z$.  Here we assume monovalent microions ($z=1$) 
dispersed in water at temperature $T=293$ K, corresponding to a Bjerrum length 
$\lambda_B\equiv e^2/(\epsilon k_BT)=0.72$ nm.  Assuming $N_m$ macroions and 
$N_s$ pairs of dissociated salt ions, we have $N_+=(Z/z)N_m+N_s$ positive and 
$N_-=N_s$ negative microions.

\subsection{Coarse-Grained One-Component Model}\label{OCM}
\vspace*{-0.2cm}
Long-ranged Coulomb interactions and high charge asymmetries render large-scale
simulations of the primitive model computationally challenging.  The model can 
be further simplified, however, by averaging over microion degrees of freedom 
to map the macroion-microion mixture onto a coarse-grained one-component model 
governed by effective electrostatic interactions~\cite{rowlinson84}.  
The mapping acts on the partition function,
\begin{equation}
{\cal Z}=\langle\langle\exp(-\beta H)\rangle_{\mu}\rangle_m, 
\label{part1}
\end{equation}
where $H$ is the total Hamiltonian, $\beta\equiv 1/k_BT$, and angular brackets 
denote traces over microion ($\mu$) and macroion ($m$) degrees of freedom.
The Hamiltonian naturally decomposes, according to $H=H_m+H_{\mu}+H_{m\mu}$, 
into a bare macroion Hamiltonian $H_m$, a microion Hamiltonian $H_{\mu}$, 
and a macroion-microion interaction energy $H_{m\mu}$.  For a chemically 
closed suspension, which exchanges no particles with its surroundings, a 
canonical trace over only microion coordinates yields the canonical partition 
function
\begin{equation}
{\cal Z}=\la\exp(-\beta H_{\rm eff})\ra_m,
\label{part2}
\end{equation}
where $H_{\rm eff}=H_m+F_{\mu}$ is the effective one-component Hamiltonian and
\begin{equation}
F_{\mu}=-k_BT\ln\la\exp\left[-\beta(H_{\mu}+H_{m\mu})\right]\ra_{\mu}
\label{Fmu1}
\end{equation}
is the Helmholtz free energy of a microion gas in the midst of fixed macroions. 
Equations (\ref{part2}) and (\ref{Fmu1}) provide a formal basis for 
approximating effective electrostatic interactions and simulating the 
effective one-component model of charged colloids.

\section{Theory}\label{Theory}

\subsection{Linear-Response Theory}\label{Response}
\vspace*{-0.2cm}
Statistical mechanical descriptions of effective electrostatic interactions, 
including density-functional~\cite{vRH97,vRDH99,vRE99,zoetekouw_pre06}, 
extended Debye-H\"uckel~\cite{warren00,chan85,chan01}, and 
response~\cite{silbert91,denton99,denton00,denton04} theories, typically
invoke linearization and mean-field approximations for the microion free energy
$F_{\mu}$ [Eq.~(\ref{Fmu1})].  Response theory describes the perturbation of 
the microion densities by the ``external" macroion electrostatic potential.  
Taking as the unperturbed reference system a uniform gas of microions in
the free volume outside the macroion hard cores, the microion free energy 
can be expressed as
\vspace*{-0.4cm}
\begin{equation}
F_{\mu}=F_{\rm plasma}+\int_0^1{\rm d}\lambda\, \la H_{m\mu}\ra_{\lambda}-E_b,
\label{Fmu2}
\end{equation}
where $F_{\rm plasma}$
is the free energy of a uniform plasma of microions in a charge-neutralizing
background of energy $E_b$, the charging parameter $\lambda$ tunes the 
macroion charge (and microion response) from zero to maximum, and 
$\la~\ra_{\lambda}$ represents an average with respect to an ensemble of 
macroions charged to a fraction $\lambda$ of their full charge.  
For weakly correlated microions, the plasma free energy has the ideal-gas form, 
\begin{equation}
\beta F_{\rm plasma}=N_+[\ln(n_+\Lambda_{\mu}^3)-1]+
N_-[\ln(n_-\Lambda_{\mu}^3)-1],
\label{Fplasma}
\end{equation}
where $n_{\pm}=N_{\pm}/[V(1-\eta)]$ are the average microion number densities,
$\eta=(4\pi/3)n_m a^3$ is the volume fraction of the macroions with number 
density $n_m=N_m/V$, and $\Lambda_{\mu}$ is the microion thermal wavelength.

The linear-response approximation expands the microion number densities in 
functional Taylor series in powers of the macroion external potential, 
truncates the expansions at linear order, and neglects microion correlations 
by assuming mean-field response 
functions~\cite{silbert91,denton99,denton00,denton04}.
The resulting internal potential energy,
\begin{equation}
U=E_{\rm vol}(N_m,N_s,V)+U_{\rm pair}(\{{\bf r}\};N_m,N_s,V),
\label{U}
\end{equation}
separates into a one-body volume energy $E_{\rm vol}$, which is independent 
of macroion coordinates, and a pair potential energy $U_{\rm pair}$, which 
depends on the macroion coordinates $\{{\bf r}\}$.
The volume energy, originating from the microion entropy and macroion-microion 
interaction energy, is given by
\begin{eqnarray}
\beta E_{\rm vol}&=&\beta F_{\rm plasma}-N_m\left(\frac{Z}{z}\right)^2
\frac{\lambda_B}{2}\frac{\kappa}{1+\kappa a} \nonumber \\
&-&N_m\frac{Z}{2}\frac{n_+-n_-}{n_{\mu}},
\label{Evol}
\end{eqnarray}
where $\kappa=\sqrt{4\pi\lambda_B z^2 n_{\mu}}$ is the Debye screening constant
(inverse screening length), a function of the total microion density, 
$n_{\mu}=n_++n_-$.  The pair potential energy, 
\begin{equation}
U_{\rm pair}=\frac{1}{2}\sum_{i\neq j=1}^{N_m}v_{\rm eff}(|{\bf r}_i-{\bf r}_j|),
\label{Upair}
\end{equation}
is a sum of hard-sphere-repulsive-Yukawa (screened-Coulomb) effective 
pair potentials,
\begin{equation}
v_{\rm eff}(r)=\left\{ \begin{array} {l@{\quad}l}
\frac{\displaystyle Z^2e^2}{\displaystyle \epsilon}\left(
\frac{\displaystyle \exp({\kappa a})}{\displaystyle 1+\kappa a}\right)^2~
\frac{\displaystyle \exp({-\kappa r})}{\displaystyle r}, & r\ge 2a, \\
\infty, & r<2a.
\label{veffr}
\end{array} \right.
\end{equation}
The effective pair potential, a product of microion screening of the bare 
macroion-macroion Coulomb interactions, has the long-range form of the DLVO 
potential~\cite{DL,VO}, but with a density-dependent screening constant.  
The constraint of electroneutrality ties average macroion and microion number 
densities via $Zn_m/(1-\eta)=z(n_+-n_-)$, rendering the effective interactions 
dependent on the average densities of both macroions and salt ion pairs, 
$n_s=N_s/[V(1-\eta)]$.  Equations (\ref{Fplasma})-(\ref{veffr}) summarize 
the effective interactions that we input to theory and simulations of the 
one-component model.

\subsection{Variational Free Energy Theory}\label{Variational}

At constant particle numbers, volume, and temperature, the Helmholtz free energy
$F$ is a minimum with respect to variations in $N_m$, $N_{\pm}$, $V$, and $T$
at thermodynamic equilibrium.  The electroneutrality constraint requires that 
ion exchange between phases occurs only in charge-neutral units, allowing the 
free energy to be regarded as a function of the number of salt ion pairs $N_s$, 
rather than of $N_+$ and $N_-$ separately.  
Within the one-component model, the free energy separates, according to
\begin{eqnarray}
F(N_m,N_s,V)&=&F_{\rm id}(N_m,V)+F_{\rm ex}(N_m,N_s,V) \nonumber \\
&+&E_{\rm vol}(N_m,N_s,V),
\label{Ftot}
\end{eqnarray}
where $F_{\rm id}=N_mk_BT[\ln(n_m\Lambda^3)-1]$ is the free energy of an 
ideal (noninteracting) gas of macroions of thermal wavelength $\Lambda$, 
and $F_{\rm ex}$ is the excess free energy due to effective pair interactions
[Eq.~(\ref{veffr})].  

A variational approximation~\cite{vRH97,vRDH99,vRE99,denton06} based on 
first-order thermodynamic perturbation theory with a hard-sphere (HS) 
reference system~\cite{HM} gives the excess free energy density as 
\begin{eqnarray}
&&f_{\rm ex}(n_m,n_s)=\min_{(d)}\left\{f_{\rm HS}(n_m,n_s;d)+2\pi n_m^2
\phantom{\int}
\right.
\nonumber \\
&\times&\left. \int_d^{\infty}{\rm d}r\, r^2 
g_{\rm HS}(r,n_m;d) v_{\rm eff}(r,n_m,n_s)\right\},
\label{fex}
\end{eqnarray}
where the effective HS diameter $d$ is the variational parameter and 
$f_{\rm HS}(n_m,n_s;d)$ and $g_{\rm HS}(r,n_m;d)$ are, respectively, 
the excess free energy density and (radial) pair distribution function 
of the HS fluid, computed here from the near-exact Carnahan-Starling and 
Verlet-Weis expressions~\cite{HM}.  According to the Gibbs-Bogoliubov 
inequality~\cite{HM}, minimization of $f_{\rm ex}$ with respect to $d$ 
yields a least upper bound to the free energy.
From the variational approximation for the total free energy [Eqs.~(\ref{Ftot})
and (\ref{fex})], the fluid branch of the phase diagram can be computed by 
performing a common-tangent construction on the curve of free energy density 
$f=F/V$ vs. macroion number density $n_m$ at fixed salt chemical potential, 
imposing equality of the pressure, 
$p=n_m(\partial f/\partial n_m)_{N_s/N_m}-f$, and of the macroion and salt 
chemical potentials, $\mu_m=(\partial f/\partial n_m)_{n_s}$ and 
$\mu_s=(\partial f/\partial n_s)_{n_m}$, in coexisting phases.

\section{Monte Carlo Simulations}\label{Simulations}

The effective interactions described above, which were used in previous 
variational theory calculations for the one-component model~\cite{denton06},
are here input into simulations of the same model to test the accuracy of 
the variational approximation and its predictions for thermodynamic behavior.  
The Gibbs ensemble Monte Carlo (GEMC) 
method~\cite{panagiotopoulos00,panagiotopoulos87,panagiotopoulos88,
panagiotopoulos89,panagiotopoulos95} is an efficient means of simulating 
two-phase fluid coexistence that obviates the need to model interfaces.  
Each phase is represented by its own simulation box,
with fluctuating macroion numbers $N_{mi}$ and volumes $V_i$ ($i=1,2$).
In the constant-$NVT$ implementation, the total macroion number, 
$N_m=N_{m1}+N_{m2}$, total volume, $V=V_1+V_2$, and 
temperature $T$ all remain fixed.  We further fix the total number of 
(implicit) salt ion pairs, $N_s=N_{s1}+N_{s2}$, while performing virtual 
exchanges between boxes.  Although the GEMC method has been previously 
applied to fluids with density-dependent pair potentials~\cite{smit92}, 
it has not yet, to our knowledge, been adapted to charged systems whose
effective interactions include both a pair potential and volume energy. 

The conventional GEMC algorithm~\cite{panagiotopoulos00,panagiotopoulos87,
panagiotopoulos88,panagiotopoulos89,panagiotopoulos95} involves three types 
of random trial move: 
(1) displacements of particles (macroions) within each box to ensure 
thermal equilibrium of each phase; 
(2) volume exchanges between the two boxes to ensure mechanical equilibrium, 
characterized by equality of pressures; and 
(3) macroion transfers between the two boxes to ensure chemical equilibrium 
with respect to macroion exchange, characterized by equality of macroion 
chemical potentials.  The acceptance probability $P_{\rm move}$ for any 
trial move from an old ($o$) to a new ($n$) state can be derived from 
the Metropolis condition~\cite{metropolis53,frenkel01,allen},
\begin{equation}
P_{\rm move}=\min\left\{\frac{{\cal P}(n)}{{\cal P}(o)},1\right\}, 
\label{Pmove}
\end{equation}
where the Gibbs ensemble probability density~\cite{frenkel01} is given by
\begin{eqnarray}
{\cal P}&=&\frac{1}{N_{m1}!N_{m2}!}\left(\frac{V_1}{\Lambda^3}\right)^{N_{m1}}
\left(\frac{V_2}{\Lambda^3}\right)^{N_{m2}} \nonumber \\[1ex]
&\times&\exp[-\beta U(\{{\bf s}\};n_m,n_s)]
\label{P}
\end{eqnarray}
and $\{{\bf s}\}$ denotes the macroion coordinates scaled by their respective 
box lengths.  Although the salt ion coordinates do not explicitly appear 
in Eq.~(\ref{P}), the potential energy $U$ [Eq.~(\ref{U})] implicitly depends 
on the average salt (and macroion) densities in the two boxes. 

From Eqs.~(\ref{Pmove}) and (\ref{P}), trial displacements are accepted 
with probability 
\begin{equation}
P_{\rm disp}=\min\left\{\exp(-\beta\Delta U),1\right\},
\label{acc-disp}
\end{equation}
where $\Delta U=U(n)-U(o)$ is the change in total potential energy 
between the new and old states.  Note that for internal displacements,
which do not affect the volume energy, $\Delta U=\Delta U_{\rm pair}$
[Eq.~(\ref{Upair})].  For all other moves, however, the change in total 
potential energy also includes a change in volume energy:
$\Delta U=\Delta E_{\rm vol}+\Delta U_{\rm pair}$.

A trial exchange of volume $\Delta V$ from box 1 to box 2 
($V_1\to V_1-\Delta V$, $V_2\to V_2+\Delta V$) is achieved by uniformly 
rescaling all macroion coordinates.  In practice, it proves more efficient 
to vary $\ln(V_1/V_2)$, with an acceptance probability~\cite{frenkel01}
%\begin{equation}
%P_{\rm vol}=\min\left\{\left(\frac{V_1-\Delta V}{V_1}\right)^{N_{m1}+1} 
%\left(\frac{V_2+\Delta V}{V_2}\right)^{N_{m2}+1} \break
%\times\exp(-\beta\Delta U),1\right\}.
%\label{acc-vol}
%\end{equation}
\begin{eqnarray}
P_{\rm vol}&=&\min\left\{\left(\frac{V_1-\Delta V}{V_1}\right)^{N_{m1}+1} 
\left(\frac{V_2+\Delta V}{V_2}\right)^{N_{m2}+1} \right.
\nonumber \\
&\times& \left. \phantom{\int}\hspace*{-0.4cm} \exp(-\beta\Delta U),1\right\}.
\label{acc-vol}
\end{eqnarray}
Transfer of a macroion from box 1 to box 2 ($N_{m1}\to N_{m1}-1$, 
$N_{m2}\to N_{m2}+1$) is accepted with probability~\cite{panagiotopoulos88}
\begin{equation}
P_{\rm trans}=\min\left\{\frac{N_{m1}}{N_{m2}+1}~\frac{V_2}{V_1}
~\exp(-\beta\Delta U),1 \right\}.
\label{acc-transfer-macroion}
\end{equation}
Note that $\Delta U$ in Eqs.~(\ref{acc-vol}) and (\ref{acc-transfer-macroion}),
represents the change in total potential energy of the two boxes combined,
since exchanges of volume or macroions alter the average macroion density 
($n_{mi}=N_{mi}/V_i$), and thus the volume energy [Eq.~(\ref{Evol})] and
pair potential [Eq.~(\ref{veffr})], in each box.

In addition to the conventional GEMC moves, we introduce a new trial move: 
transfer of salt between the two boxes, required to ensure chemical equilibrium 
with respect to salt exchange between coexisting phases, characterized by 
equality of salt chemical potentials.  Since the salt is modeled here only 
implicitly, virtual transfers involve simply changing the average 
salt density of each box, with acceptance probability 
\begin{equation}
P_{\rm salt}=\min\left\{\exp(-\beta\Delta U),1 \right\},
\label{acc-transfer-salt}
\end{equation}
where $\Delta U$ is the change in total potential energy of both boxes.  
We stress that exchanges of average salt density affect both the pair 
potential and the volume energy in each box.  The absence of combinatorial 
and phase-space prefactors in Eq.~(\ref{acc-transfer-salt}) follows from 
implicit modeling of salt ions.  In practice, a transfer of $\Delta N_s$ 
salt ion pairs from box 1 to box 2 ($N_{s1}\to N_{s1}-\Delta N_s$, 
$N_{s2}\to N_{s2}+\Delta N_s$) is realized by changing the respective salt 
densities accordingly and adjusting $\Delta N_s$ ($\ll N_{s1},N_{s2}$) 
to achieve a reasonable acceptance rate.  

Within the Gibbs ensemble, we simulated two cubic boxes subject to periodic 
boundary conditions, each box containing only macroions, but evolving 
according to effective interactions [Eqs.~(\ref{Evol}) and (\ref{veffr})]
that implicitly depend on the microion densities.  To exclude interactions 
of a particle with its own periodic images, and avoid needless computation, 
pair interactions between macroions were cut off at a distance of 
$r_c=\min\{20/\kappa,L/2\}$, i.e., the shorter of 20 screening lengths or 
half the respective box length $L$.  The effective interactions were updated 
whenever the average macroion or salt density changed.  

The simulations started from initial configurations of randomly distributed 
macroions, with equal particle numbers, volumes, and salt concentrations in 
each box.
The four types of trial move were executed in random sequence at prescribed
frequencies.  Defining a cycle as an average of $N_m$ trial displacements
(i.e., one per macroion), the other moves were attempted with relative 
frequencies per cycle of $N_m/2$ for volume exchanges, $N_m/10$ for macroion 
transfers, and $N_m/10$ for salt exchanges.  For internal displacements, 
macroions were selected at random and moved with tolerances adjusted to yield 
an acceptance rate of about 50\%.  For volume and salt exchanges, acceptance 
rates of about 10\% were achieved by adjusting the tolerances, the resulting 
salt tolerance being $\Delta N_s/N_s\simeq 10^{-3}$.  After equilibrating for 
$10^4$ cycles, we accumulated statistics for average densities, pressures,
and chemical potentials over the next $10^4$ cycles ($5\times 10^6$ 
displacements for $N_m=500$).

\section{Results and Discussion}\label{Results}

\subsection{Tests of One-Component Model \\ and Variational Theory}

To investigate thermodynamic phase behavior of charged colloids, we input 
effective electrostatic interactions (Sec.~\ref{Response}) to both variational 
theory calculations (Sec.~\ref{Variational}) and Gibbs ensemble Monte Carlo 
simulations (Sec.~\ref{Simulations}) of the coarse-grained one-component model.
The validity of the one-component model is first tested by comparing 
structural and thermodynamic properties with available data from simulations 
of the primitive model, which include explicit point counterions interacting 
via bare Coulomb potentials.

From extensive Monte Carlo (MC) simulations, Linse~\cite{linse00} has generated
a wealth of data for the (salt-free) primitive model over ranges of 
macroion valence, volume fraction, and electrostatic coupling parameter, 
$\Gamma=\lambda_B/a$.  For direct comparison, we performed simulations of the 
effective one-component model for identical parameters -- fixing the effective 
macroion valence ($Z=40$), counterion valence ($z=1$), and Bjerrum length
($\lambda_B=0.72$ nm), and varying the macroion radius $a$ -- and computed the 
macroion-macroion pair distribution function $g(r)$ and pressure $p$, as 
described in the Appendix.  For this purpose, we performed standard 
constant-$NVT$ (one-box) simulations, the volume energy then having no effect 
on the pair structure.  To obtain accurate $g(r)$ results, a system size 
of $N_m=600$ sufficed to render finite-size effects negligible.  To maintain
consistently high accuracy in the pressure, we increased the particle number to 
ensure a cut-off radius of at least 20 screening lengths for each combination 
of $\eta$ and $\Gamma$ --- ranging up to $N_m=8000$ for $\eta=0.02$, 
$\Gamma=0.0445$.

\begin{figure}[h]
\includegraphics[width=0.9\columnwidth,angle=0]{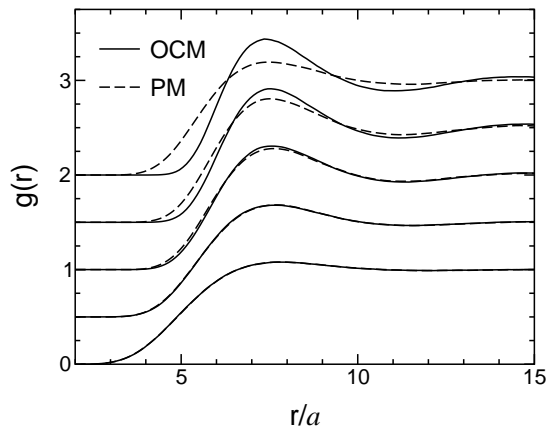}
\vspace*{-0.2cm}
\caption{
\label{gr-linse}
Macroion-macroion pair distribution function $g(r)$ vs. radial distance $r$
(units of macroion radius $a$) of salt-free suspensions computed from 
Monte Carlo simulations of the one-component model (OCM) with implicit 
counterions (solid curves) and the primitive model (PM)~\cite{linse00} 
with explicit counterions (dashed curves) for macroion valence $Z=40$, 
volume fraction $\eta=0.01$, and electrostatic coupling parameters 
$\Gamma=\lambda_B/a=$ 0.0222, 0.0445, 0.0889, 0.1779, 0.3558 (bottom to top).
For clarity, curves are vertically offset in steps of 0.5.}
\end{figure}

\begin{figure}[h]
\includegraphics[width=0.9\columnwidth]{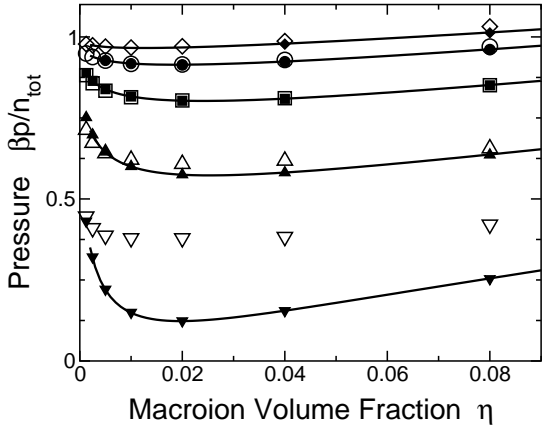}
\vspace*{-0.2cm}
\caption{
\label{p-linse}
Reduced pressure $\beta p/n_{\rm tot}$ vs. macroion volume fraction $\eta$, 
where $n_{\rm tot}=(Z+1)n_m$, for salt-free suspensions computed from 
Monte Carlo simulations of the effective one-component model with implicit 
counterions (solid symbols) and the primitive model~\cite{linse00} with 
explicit counterions (open symbols) for macroion valence $Z=40$ and 
several electrostatic coupling parameters $\Gamma=\lambda_B/a$.
Simulation error bars are smaller than symbol sizes.
Also shown are corresponding predictions of variational theory (curves).
From top to bottom, $\Gamma=0.0222$, 0.0445, 0.0889, 0.1779, 0.3558.}
\end{figure}

Figures~\ref{gr-linse} and \ref{p-linse} compare numerical results of our 
simulations of the one-component model with Linse's simulations of the 
primitive model~\cite{linse00}.  At relatively low electrostatic couplings 
($\Gamma\leq 0.1779$, $Z\lambda_B/a\leq 7$), our results for the pair 
distribution function and pressure closely match the corresponding primitive 
model data (Fig.~5 (a) and Table III of ref.~\cite{linse00}, after minor 
corrections~\cite{note1}).  It should be noted that good agreement at higher 
volume fractions is achieved only when the excluded-volume factor of 
$1/(1-\eta)$ is consistently included in the effective interactions.  
These comparisons demonstrate the accuracy of the one-component model 
with linearized effective interactions for moderately coupled systems.  
Figure~\ref{p-linse} also presents predictions for the pressure from our 
variational theory calculations.  The near-perfect alignment of theory and 
simulations of the one-component model validates the variational approximation 
over the parameter ranges studied. 
\vspace*{0.7cm}

At higher electrostatic couplings ($Z\lambda_B/a>7$), typical of highly charged 
latex particles and ionic surfactant micelles, significant deviations between 
the one-component and primitive models abruptly emerge ($\Gamma=0.3558$ in 
Figs.~\ref{gr-linse} and \ref{p-linse}).  The discrepancies in this relatively 
strong-coupling regime can be traced to renormalization of the effective 
macroion charge through strong association of counterions, a nonlinear effect 
neglected in the present version of the model.  Preliminary 
investigations~\cite{denton-lu-cr} indicate, however, that the 
deviations can be substantially reduced by consistently building into the 
one-component model a renormalized effective charge.  These results establish 
a threshold of $Z\lambda_B/a\simeq 7$ for significant charge renormalization 
within linear-response theory.

\begin{figure}[h]
\includegraphics[width=0.9\columnwidth]{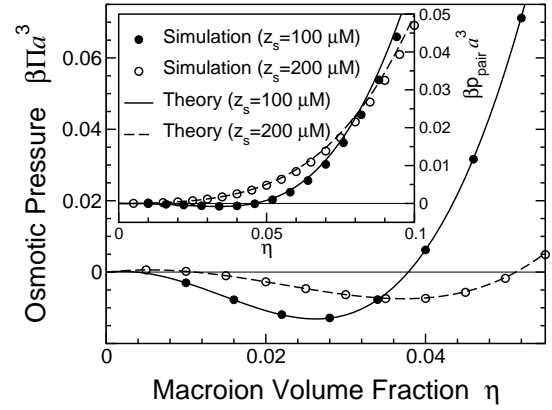}
%\vspace*{-0.2cm}
\caption{
\label{press-fig}
Reduced osmotic pressure $\beta\Pi a^3$ vs. volume fraction $\eta$ computed 
from Monte Carlo simulations and variational theory~\cite{denton06} 
of the one-component model for Bjerrum length $\lambda_B=0.72$ nm, 
macroion radius $a=50$ nm, valence $Z=500$, and salt fugacities $z_s=100$ 
and 200 $\mu$M.  Changes of curvature reflect phase instability.  
Inset: Pair potential contribution to total pressure.
Simulation error bars are smaller than symbol sizes.}
\end{figure}

To test the variational free energy theory at higher charge asymmetries and 
nonzero salt concentrations, we compare predictions for the osmotic pressure 
(equation of state) with results from our GEMC simulations.  The osmotic 
pressure, $\Pi=p-2n_rk_BT$, is here defined as the total pressure of the 
suspension less that of a (virtual) ideal-gas salt reservoir 
of ion pair density $n_r$ at the same salt chemical potential:
$\mu_s=k_BT\ln(2n_r\Lambda_{\mu}^3)$.  Figure~\ref{press-fig} shows sample 
results for the equation of state at fixed salt chemical potential or, 
equivalently, salt fugacity, $z_s=\exp(\beta\mu_s/2)$, from simulations 
at the salt concentrations predicted by theory for each volume fraction.  
Since theory and simulation assume identical effective interactions, 
the comparisons directly probe the excess free energy approximation 
[Eq.~(\ref{fex})] and corresponding pair potential contribution to the total 
pressure (inset to Fig.~\ref{press-fig}).  The predictions are in excellent 
agreement with simulation over a wide range of volume fractions, further 
validating the variational approximation and providing a consistency check
on our calculations.  As an independent check, our methods accurately reproduce 
pressures computed from MC simulations of the 
hard-sphere-repulsive-Yukawa pair potential fluid~\cite{cochran-chiew04}. 

The appearance in Fig.~\ref{press-fig} of a van der Waals loop in the pressure 
signals a spinodal instability and separation into macroion-rich (liquid) and 
macroion-poor (vapor) phases.  We stress, however, that currently available 
data from primitive model simulations can test the effective one-component 
model and linearized effective interactions only for salt-free suspensions at 
relatively low charge asymmetries, where instabilities with respect to phase 
separation have not been predicted.  
Furthermore, the macroion aggregation observed in ref.~\cite{linse00} in the 
strong-coupling regime is likely driven by microion correlations, which are 
neglected in the mean-field effective interactions assumed here.  While 
further tests of the one-component model are needed, the close agreement 
for parameters accessible to primitive model simulations motivates proceeding
to consider phase behavior.

\subsection{Phase Behavior}

\begin{figure}[h]
\includegraphics[width=\columnwidth]{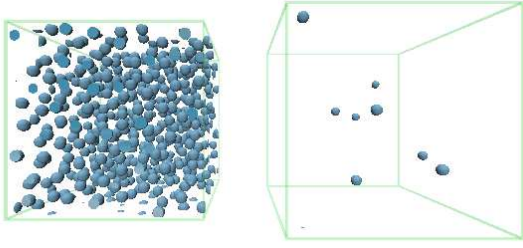}
\vspace*{-0.2cm}
\caption{Snapshot from Gibbs ensemble Monte Carlo simulation,
showing the two boxes after separation into macroion-rich and macroion-poor 
phases.  Spheres depict macroions in the effective one-component model.
}
\label{snapshots}
\end{figure}

To systematically map out the fluid binodal, we performed a series of GEMC 
simulations over ranges of volume fraction and salt concentration for selected 
macroion radii and valences: ($a=10$ nm, $Z=150$) and ($a=50$ nm, $Z=500$).
Initially uniform systems of $N_m=500$ particles (two-box total), in
thermodynamic states ($\eta$, $c_s$) within the predicted unstable 
region~\cite{denton06}, spontaneously separated into two phases, each phase 
occupying one of the boxes, which differed in average macroion and salt 
densities.  In contrast, systems at state points outside of the unstable 
region remained uniform.  A visual impression of the phase separation is 
provided by the simulation snapshot in Fig.~\ref{snapshots}.

To identify the structure of the coexisting phases, we performed constant-$NVT$
(one-box) simulations, at identical state points, for particle numbers commensurate 
with likely crystal structures: {\it fcc} ($N_m=500$) and {\it bcc} ($N_m=432$).
Initializing the particles on the sites of the respective lattice, we computed 
the equilibrium pair distribution function and observed typical fluid-like 
structure, indicating melting of the initial crystal.  Upon increasing the 
volume fraction, we observed, at state points well outside the fluid binodal, 
an abrupt sharpening of the peaks of $g(r)$, reflecting crystallization.  
These observations are consistent with a simple hard-sphere freezing criterion,
$\eta(d/2a)^3\simeq 0.49$, which approximates the macroions as hard spheres of 
effective diameter $d$ [from Eq.~(\ref{fex})] and locates the coexistence 
densities within the fluid regime.

\begin{figure}[h]
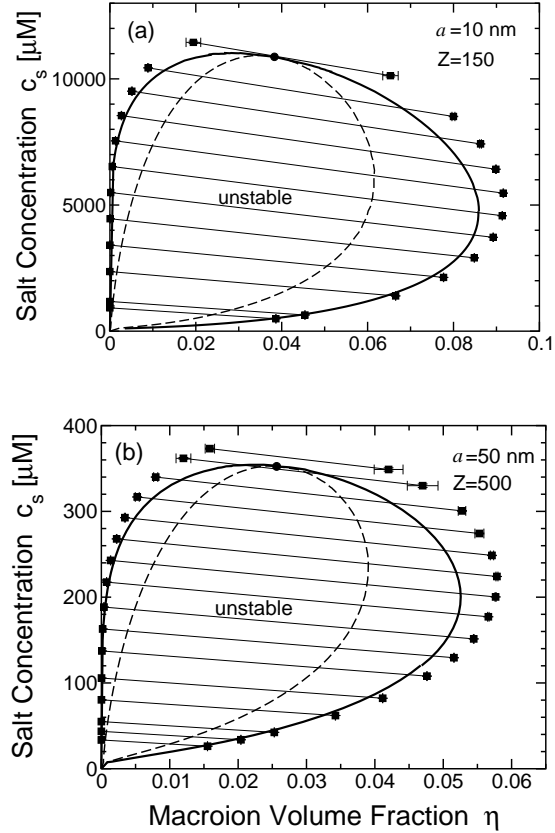

\includegraphics[width=0.9\columnwidth]{s20z150phase.new.eps} 
\\[2ex]
\includegraphics[width=0.9\columnwidth]{s100z500phase.new.eps}
\vspace*{-0.2cm}
\caption{
\label{pdlv100}
Phase diagrams of aqueous charged colloids, showing fluid binodal computed 
from Gibbs ensemble Monte Carlo simulations (squares) and variational 
theory~\cite{denton06} (solid curves) for the effective one-component model 
at various combinations of macroion radius $a$ and valence $Z$.  
Tie lines join corresponding points on the colloid-rich and colloid-poor 
branches of the binodal.  Also shown are predictions of variational theory 
for the critical point (circles) and spinodal (dashed curves).}
\end{figure}

The resulting phase diagrams are presented in Fig.~\ref{pdlv100}, alongside
predictions of variational theory~\cite{denton06}, where tie lines joining 
corresponding points on the macroion-rich and macroion-poor binodal branches 
parallel those predicted by theory.  Each pair of points on the binodal
was produced by averaging over 10 independent runs, which differed only in 
the random number seed used for trial moves.  Reported error bars represent 
statistical uncertainties of one standard deviation, computed from fluctuations
in average densities among the 10 runs.  Resolution near the critical point 
is blurred by density fluctuations and phase switching between boxes -- a known
limitation of the Gibbs ensemble method~\cite{frenkel01}.  For simplicity, we
discarded runs in which the phases switched boxes, a rare occurrence away from 
the critical point.  Considering the sensitivity of the coexistence analysis to 
slight deviations in free energy, the quantitative agreement between theory 
and simulation attests to the accuracy of the variational approximation.
\\[1ex]

\begin{figure}[h!]
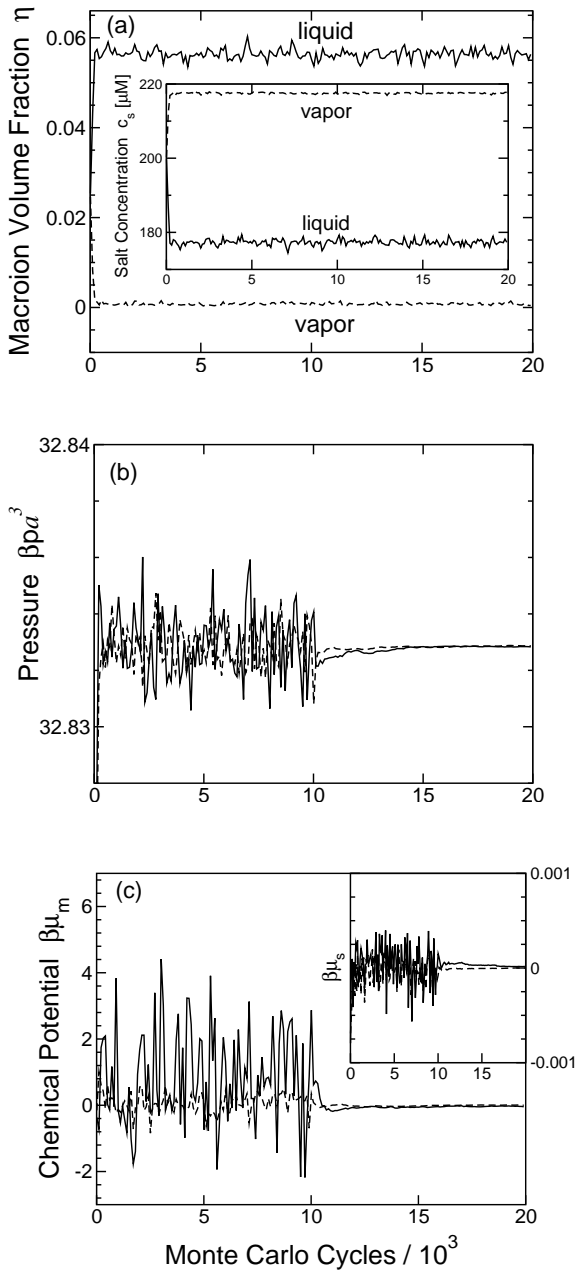

\includegraphics[width=0.9\columnwidth]
{eta_vs_MC_s100z500cs200e0025MC20000.eps} \\[1.2ex]
\includegraphics[width=0.9\columnwidth]
{p_vs_MC_s100z500cs200e0025MC20000.eps} \\[1.2ex]
\hspace*{0.8cm}\includegraphics[width=0.9\columnwidth]
{cp_vs_MC_s100z500cs200e0025MC20000.eps}
\vspace*{-0.2cm}
\caption{
\label{diagnostics}
Evolution of diagnostic properties during a sample Gibbs ensemble Monte Carlo 
run for macroion radius $a=50$ nm, valence $Z=500$, total volume fraction 
$\eta=0.025$, and total salt concentration $c_s=200$ $\mu$M.  
Solid and dashed curves represent the two boxes.
(a) volume fractions and (inset) salt concentrations;
(b) total pressure;
(c) macroion and (inset) salt chemical potentials,
shifted to zero average.
Instantaneous values are plotted during equilibration (first $10^4$ cycles)
and cumulative values thereafter.
}
\end{figure}

Diagnostic variables were monitored during the simulations and evolved as 
typified by Fig.~\ref{diagnostics}, which tracks the volume fractions, salt 
concentrations, pressures, and chemical potentials in each box vs. number of 
MC cycles for one sample run.  Bifurcation of volume fractions and salt 
concentrations in the two boxes [Fig.~\ref{diagnostics} (a)] signals 
phase separation, while convergence of pressures and chemical potentials 
[Fig.~\ref{diagnostics} (b) and (c)] confirms equilibration.  Several runs 
for larger systems (up to 1000 macroions) were performed to establish the 
insignificance of finite-size effects.  
Compared with large-scale simulations of the primitive model, our simulations 
require relatively modest computing resources, each run typically consuming 
50-90 CPU hours on a single PC (Intel Xeon-HT processor).

It should be emphasized that the observed phase separation, although perhaps 
surprising in the face of purely repulsive pair interactions, is driven by 
the state dependence of the volume energy in the one-component model of 
deionized suspensions.  It is also important to note that the macroion valences
and electrostatic couplings represented in Fig.~\ref{pdlv100} were selected to 
lie just within the unstable fluid regime and correspond to 
$\Gamma=$ 0.072 and 0.014 ($Z\lambda_B/a=$ 7 and 11) in panels (a) and (b), 
respectively ({\it cf}. $Z\lambda_B/a\simeq$ 1-14 in ref.~\cite{linse00}).  
In each case, a small increase in macroion radius or decrease in valence 
stabilizes the system.  These parameters approach the threshold for charge 
renormalization estimated from our direct comparisons with primitive model 
simulations (Figs.~\ref{gr-linse} and \ref{p-linse}), albeit for lower valences.
Whether the predicted phase instability corresponds to a real phenomenon 
or is merely an artifact of the linearization 
approximation~\cite{vongrunberg01,deserno02,tamashiro03}, or the assumption of 
fixed macroion valence~\cite{levin98,levin01,levin03,levin04,zoetekouw_prl06},
remains unclear.  Preliminary explorations~\cite{denton-lu-cr}, based on a 
simple model of effective macroion valence, suggest that the instability 
survives incorporation of charge renormalization in the one-component model.  
Further studies are required, however, to resolve this important issue.

\section{Conclusions}\label{Conclusions}

In summary, we have developed a new variant of the Gibbs ensemble Monte Carlo
method to simulate a one-component model of charged colloids governed by
density-dependent effective interactions.  The effective interactions (pair 
potential and one-body volume energy) are input from linear-response theory, 
assuming a mean-field approximation for microion structure.  The simulation 
algorithm includes trial exchanges of implicit salt between the two simulation 
boxes and incorporates the volume energy into the acceptance probabilities 
for trial moves that change the average macroion or salt density.  

Comparisons with simulations of the two-component (salt-free) primitive 
model~\cite{linse00} demonstrate the validity of the one-component model 
over a wide parameter range, physically relevant to charged latex particles 
and micelles.  Results for the macroion-macroion pair distribution function 
and pressure are in close agreement with corresponding primitive model results 
for moderate electrostatic couplings.  Deviations at stronger couplings likely 
originate from nonlinear screening effects neglected in the present model.  
Our simulations also confirm the accuracy of a variational free energy
approximation~\cite{vRH97,vRDH99,vRE99,denton06}.  Further comparisons 
would help to more sharply define the limitations of the one-component model.

While the cost of primitive model simulations grows with increasing charge 
asymmetry, concentration, and electrostatic coupling, the computational effort 
required to simulate the one-component model is relatively modest and actually 
diminishes with increasing macroion valence and concentration, as decreasing 
the screening length shortens the range of effective pair interactions.  The 
one-component model thus offers insight into bulk phase behavior in parameter 
regimes that may be computationally prohibitive for more explicit models.

We have applied our new simulation method to test predictions of variational 
theory~\cite{denton06} for the phase behavior of aqueous suspensions of 
charged macroions with weakly correlated (monovalent) microions at low salt 
concentrations.  The resulting phase diagrams exhibit coexistence of
macroion-rich and macroion-poor fluid phases in excellent agreement with 
previous predictions and qualitatively consistent with observed thermodynamic
anomalies.  The phase instability predicted by theory, and now confirmed by
simulations of the same model, occurs in a parameter regime that appears to 
border the threshold for saturation of the effective macroion charge.  
Future work will address this open issue by incorporating charge 
renormalization into the one-component model~\cite{denton-lu-cr}.  
Finally, our simulation algorithm can be extended to investigate other 
phase transitions, e.g., crystallization, and adapted to model other 
soft materials, such as polyelectrolyte and ionic micellar solutions.

\vspace*{-0.5cm}
\acknowledgements
\vspace*{-0.2cm}
We thank Alexander Wagner for helpful discussions, Per Linse for sharing data 
from his primitive model simulations, and the Center for High Performance 
Computing at North Dakota State University for computing facilities.  This work 
was supported by the National Science Foundation under Grant Nos.~DMR-0204020
and EPS-0132289.

\vspace*{-0.1cm}
\appendix*
\section{Diagnostic Calculations}\label{Diagnostics}
\subsection{Pressure}\label{Pressure}
\vspace*{-0.2cm}
A diagnostic for mechanical equilibrium in the Gibbs ensemble is equality of 
pressures in the two boxes.  The total pressure naturally separates 
into three distinct contributions:
\begin{equation}
p=p_{\rm id}+p_{\rm pair}+p_{\rm vol},
\label{p}
\end{equation}
where $p_{\rm id}=\la n_m\ra k_BT$ is the ideal-gas pressure of the macroions, 
$p_{\rm pair}$ results from effective pair interactions between macroions, 
$p_{\rm vol}=-\la(\partial E_{\rm vol}/\partial V)_{N_m/N_s}\ra$ is the 
contribution from the density-dependent volume energy, and angular brackets 
denote an ensemble average over configurations in the Gibbs ensemble.  
The pair pressure is calculated on the fly within the simulations using 
the virial expression for a density-dependent pair potential~\cite{louis02}:
\begin{equation}
p_{\rm pair}=\la\frac{{\cal V}_{\rm int}}{3V}\ra
-\la\left(\frac{\partial U_{\rm pair}}{\partial V}\right)_{N_m/N_s}\ra
+p_{\rm tail},
\label{ppair}
\end{equation}
where ${\cal V}_{\rm int}$ is the internal virial, the volume derivative term 
accounts for the density dependence of the effective pair potential, 
and $p_{\rm tail}$ corrects for cutting off the long-range tail of the 
pair potential.  The internal virial is given by
\begin{equation}
{\cal V}_{\rm int}=\sum_{i=1}^{N_m}{\bf r}_i\cdot{\bf f}_i
=\sum_{i\neq j=1}^{N_m}(1+\kappa r_{ij})v_{\rm eff}(r_{ij}),
\label{virial}
\end{equation}
where ${\bf f}_i=-\nabla\sum_{j\neq i}v_{\rm eff}(r_{ij})$ is the effective 
force exerted on macroion $i$, at position ${\bf r}_i$, by all neighboring 
macroions $j$, at relative distances $r_{ij}$, within the cut-off radius $r_c$.  
The second term on the right side of Eq.~(\ref{ppair}) is 
computed via
\begin{equation}
\left(\frac{\partial U_{\rm pair}}{\partial V}\right)_{N_s/N_m}
=-\frac{n_m^2}{2N_m}\sum_{i\neq j=1}^{N_m}\left(
\frac{\partial v_{\rm eff}(r_{ij})}{\partial n_m}\right)_{N_s/N_m}
\label{dUdV}
\end{equation}
with
\begin{equation}
\left(\frac{\partial v_{\rm eff}(r)}{\partial n_m}\right)_{N_s/N_m}
=\left(\frac{\kappa^2 a^2}{1+\kappa a}-\frac{\kappa r}{2}\right)
\frac{v_{\rm eff}(r)}{n_m(1-\eta)}.
\label{dveffdnm}
\end{equation}
The tail pressure is approximated by
\begin{eqnarray}
p_{\rm tail}&=&-\frac{2\pi}{3}\la n_m^2\int_{r_c}^\infty{\rm d}r\,r^3 
v'_{\rm eff}(r)\ra
\nonumber \\
&=&\frac{2\pi}{3}\la n_m^2\left(\frac{\kappa^2 r_c^2+3\kappa r_c+3}
{\kappa^2}\right)r_c v_{\rm eff}(r_c)\ra,
\nonumber \\
\label{ptail}
\end{eqnarray}
the approximation being the neglect of pair correlations for $r>r_c$.
Finally, the volume pressure is given by
\begin{equation}
\beta p_{\rm vol}=
\la\frac{n_m}{1-\eta}\left(Z+2\frac{N_s}{N_m}-\frac{Z^2\kappa\lambda_B}
{4(1+\kappa a)^2}\right)\ra.
\label{pvol}
\end{equation}

\subsection{Chemical Potentials}\label{Chempot}

To diagnose chemical equilibrium between coexisting phases, we computed the 
chemical potentials of macroions and salt by adapting Widom's test particle 
insertion method~\cite{widom63} to the Gibbs ensemble, following
ref.~\cite{frenkel01}.  In contrast to the original method, the inserted 
ions are not treated as ghost particles in GEMC, but rather remain within 
the box into which they are successfully transferred.  
The macroion chemical potential -- the change in Helmholtz free energy 
upon adding a macroion -- can be expressed as
\begin{equation}
\mu_m=
-k_BT\ln\left(\frac{{\cal Z}_G(N_m+1,N_s,V,T)}{{\cal Z}_G(N_m,N_s,V,T)}\right),
\label{mum}
\end{equation}
where the Gibbs ensemble partition function is given by 
\begin{eqnarray}
{\cal Z}_G&=&\frac{1}{\Lambda^{3N_m}V}\sum_{N_{m1}=0}^{N_m}
\frac{1}{N_{m1}!(N_m-N_{m1})!} 
\nonumber \\
&\times&\int_0^V{\rm d}V_1\,V_1^{N_{m1}}(V-V_1)^{N_m-N_{m1}} 
\nonumber \\
&\times&\int{\rm d}{\bf s}^{N_m}\,\exp[-\beta U(\{{\bf s}\}; n_m,n_s)].
\label{part}
\end{eqnarray}
The macroion chemical potential of box 1 is thus computed from~\cite{frenkel01}
\begin{equation}
\mu_{m1}=-k_BT\ln\la\frac{V_1/\Lambda^3}{(N_{m1}+1)}
\exp(-\beta\Delta U_1^{+m})\ra,
\label{mum1}
\end{equation}
where $\Delta U_1^{+m}=U(N_{m1}+1)-U(N_{m1})$ is the change in total potential 
energy (volume energy plus pair energy) of box 1 upon insertion of a macroion.
In practice, the large change in volume energy 
$\Delta E_{\rm vol}$ resulting from a macroion insertion necessitates 
evaluating Eq.~(\ref{mum1}) by adding to and subtracting from the argument 
of the exponential a constant $c\simeq\la\Delta E_{\rm vol}\ra$:
\begin{eqnarray}
\mu_{m1}&=&-k_BT\ln\la\frac{V_1/\Lambda^3}{N_{m1}+1}
\exp[-\beta(\Delta U_1^{+m}-c~)]\ra
\nonumber \\
&+&c.
\label{mum1-1}
\end{eqnarray}

The salt chemical potential -- the change in Helmholtz free energy upon 
insertion of a salt ion pair -- can be approximated by
\begin{equation}
\mu_s=
-\frac{k_BT}{\Delta N_s}\ln\left(\frac{{\cal Z}_G(N_m,N_s+\Delta N_s,V,T)}
{{\cal Z}_G(N_m,N_s,V,T)}\right),
\label{mus}
\end{equation}
assuming that the number of exchanged salt ion pairs $\Delta N_s$ is much
less than the total number of salt ion pairs ($\Delta N_s\ll N_s$).
The salt chemical potential of box 1 is thus computed from
\begin{equation}
\mu_{s1}=-\frac{k_BT}{\Delta N_s}\ln\la\exp[-\beta(\Delta U_1^{+s}-c~)]\ra
+\frac{c}{\Delta N_s},
\label{mus1}
\end{equation}
where $\Delta U_1^{+s}=U(N_{s1}+\Delta N_s)-U(N_{s1})$ is the change in
total potential energy of box 1 upon insertion of $\Delta N_s$ salt ion 
pairs.  The absence of combinatorial and phase space factors in 
Eq.~(\ref{mus1}) follows from modeling the microions only implicitly.  
Note also that the chemical potentials are defined only to within 
arbitrary constants.
In the dilute colloid limit ($N_m\to 0$), the salt chemical potential tends 
to that of an ideal gas of salt ions
\begin{equation}
\mu_s^{(0)}=2k_BT\la\ln(n_s\Lambda_{\mu}^3)\ra
\label{musvol}
\end{equation}
and the macroion chemical potential reduces to
\begin{eqnarray}
\mu_m^{(0)}&=&k_BT\la-\ln\left(\frac{V}{\Lambda^3}\right)+
Z\ln(n_s\Lambda_{\mu}^3) \right.
\nonumber \\
&-&\left.\phantom{\int}\hspace*{-0.4cm}\left(\frac{Z}{z}\right)^2
\frac{\kappa\lambda_B}{2(1+\kappa a)}+\frac{8\pi}{3}n_sa^3\ra,
\label{mumvol}
\end{eqnarray}
where the terms on the right side are derived (left to right) from 
the macroion entropy, microion entropy, macroion-counterion interaction, 
and macroion excluded volume.  These analytical results [Eqs.~(\ref{musvol})
and (\ref{mumvol})] provide a check on the numerical results in the limit 
in which one box becomes depleted of macroions.

\subsection{Pair Distribution Function}\label{pdf}

The structure of the suspension is characterized by the pair distribution 
functions~\cite{HM}.  The macroion-macroion pair distribution function $g(r)$ 
-- the only one accessible in the one-component model -- is defined such that 
$4\pi r^2g(r){\rm d}r$ equals the average number of macroions in a spherical 
shell of radius $r$ and thickness ${\rm d}r$ centered on a macroion.
For a given configuration, each particle is regarded, in turn, as the central 
particle.  Neighboring particles are then assigned, according to their radial 
distance $r$ from the central particle, to concentric spherical shells (bins) 
of thickness $\Delta r=0.1 a$.  After equilibration, $g(r)$ is computed, 
in the range $2a<r<L/2$, by accumulating the numbers of particles in 
radial bins and averaging over all configurations.  The resulting
distributions are finally smoothed by averaging each bin with its
immediate neighboring bins. 
\\

%%%%%%%%%%%%%%%%%%%%%%%%%%%%%%%%%%%%%%%%%%%%%%%%%%%%%%%%%%%%%%%%%%%%%%%%%%%
%                           REFERENCES
%%%%%%%%%%%%%%%%%%%%%%%%%%%%%%%%%%%%%%%%%%%%%%%%%%%%%%%%%%%%%%%%%%%%%%%%%%%

\newpage

%%%%%%%%%%%%%%%%%%%%%%%%%%%%%%%%%%%%%%%%%%%%%%%%%%%%%%%%%%%%%%%%%%%%%%%%%%%
%%%%%%%%                       FIGURES
%%%%%%%%%%%%%%%%%%%%%%%%%%%%%%%%%%%%%%%%%%%%%%%%%%%%%%%%%%%%%%%%%%%%%%%%%%%

%  use \protect\command{}  if you have to use a command \command{}
%  which has an argument

% in order to put the figures into the text you have to activate
% the line with ``\input{psfig}'' as well

\end{document}